\documentstyle[prl,aps,epsf]{revtex}

\newcommand{\newc}{\newcommand}
\newc{\beq}    {\begin{equation}}
\newc{\eeq}    {\end{equation}}
\newc{\beqa}    {\begin{eqnarray}}
\newc{\eeqa}    {\end{eqnarray}}

\begin{document}

\draft
\twocolumn[\hsize\textwidth\columnwidth\hsize\csname
@twocolumnfalse\endcsname

\title{ Qubit geometry and conformal mapping}
\author{ Jae-weon Lee\cite{leejw}, Chang Ho Kim, and Eok Kyun Lee  \\}
\address{
 Department of Chemistry,  School of Molecular Science (BK 21),
\\Korea Advanced   Institute of Science and Technology,  Taejon
 305-701, Korea.}

\author{ Jaewan Kim and Soonchil Lee}
\address{
Department of Physics, Korea Advanced Institute of Science and
Technology, Taejon 305-701, Korea}

\maketitle

\begin{abstract}
Identifying the Bloch sphere
with the Riemann sphere(the extended complex plane), we obtain
relations between
 single qubit unitary
operations and M\"{o}bius transformations on the extended complex plane.

\end{abstract}

]


In the past ten years, there has been a growing interest in the
theory of quantum computation and quantum information\cite{summary}.
Due to the quantum parallelism, quantum computers have the
potential to outperform the classical computers.
The quantum computers are based on two-level quantum systems, which
are called qubits\cite{benioff}.
In this paper, we  will investigate a representation  of
a single qubit operation on a complex plane.
The idea that a state in a two-level  quantum system can be
described by a complex number in the complex plane
is not new\cite{penrose,urbantke,mosser,havel}.
The relationship between conformal mappings and spinor transforms
is extensively studied by Rindler and Penrose\cite{penrose}.
Urbantke\cite{urbantke}
considered a stereographic projection  of two-level quantum
 system onto a complex plane, and
their time evolution on the spheres.
Recently, his work is extended by Mosser and Danoloff \cite{mosser},
who consider a  $S^7$ Hopf fibration
using  quaternions instead of complex numbers for two qubits\cite{hopf}.
The main purpose of our study is to reveal the relation between
the single qubit unitary operations and conformal mappings
which are very useful calculation tools
in theoretical  physics, such as the string theory\cite{string}

Let us begin with a brief review of the stereographic projection
with the Bloch sphere.
The Bloch sphere representation  is well known in
nuclear magnetic resonance, quantum optics and quantum
computing\cite{bloch}.
 In the computational basis, the general one qubit state
is denoted by
$|\psi \rangle= a |0 \rangle   + b|1 \rangle \equiv (a,b)^T$ , where $|a|^2
+|b|^2=1$.
Up to a  global phase factor
this state can be  written in the form
\beq
 |\psi \rangle= cos \frac{\phi}{2} |0 \rangle   + e^{-i
\theta} sin \frac{\phi}{2} |1 \rangle,
\label{definition}
\eeq
where  angles $0\leq \phi \leq \pi$ and $0\leq\theta < 2\pi$ define a point (Bloch vector) on
 the Bloch sphere  as shown in Fig.1.
The Riemann sphere, also called  the extended
complex plane,  is a  1-point compactification
of the complex plane, $C \cup \infty$, denoted as  $\tilde{C}$.
Fig.1 also illustrates the definition of the
stereographic projection, which maps a point
 $q=(x_1,x_2,x_3)$ on the unit sphere $S^2$ (the Bloch
 sphere)  to a point
 $z=x+iy$ on $\tilde{C}$.
 Here $x,y$ axes are equal to $x_1,x_2$ axes , respectively.
The  line passing  through  the north pole
$N=(0,0,1)$ and  the point $q$ on $S^2$
intersects   $\tilde{C}$  at the point $z$.
 By identifying the Bloch sphere with the Riemann sphere,
 one can define a
projection  $P: {\cal{H}} \rightarrow \tilde{C} $ from the
two-dimensional projective
Hilbert space   $\cal{H}$ to the
 extended complex plane ($\tilde{C}$).

\begin{figure}[Fig1]
\epsfysize=6cm \epsfbox{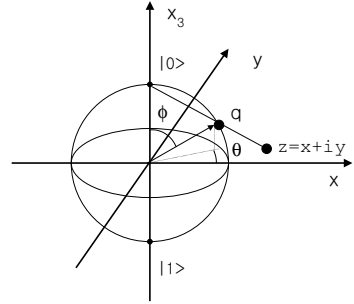} \caption[Fig1]{\label{fig1}
Stereographic projection of the Bloch sphere onto
the extended complex plane.}

\end{figure}

Then,  $P$ maps
$|\psi\rangle =(a,b)^T$  to a complex number
\beqa
z&=&P(|\psi \rangle )=
\frac{x_1}{1-x_3} +i \frac{x_2}{1-x_3}
\\ \nonumber
&=&e^{i \theta} cot \frac{\phi}{2}=\frac{a}{b}.
\label{z1}
\eeqa
Here  $a$ and $b$  are complex homogeneous
coordinates for the point $z$ relating the spinor space
 to the one-dimensional complex projective
space $CP^1$
as described in  Ref.3.
Thus, for example, $P(|0\rangle )=\infty$,  $P(|1\rangle )=0$ and
$P( [|0\rangle\pm |1\rangle]/\sqrt{2} )=\pm 1$.
If $|\psi_1 \rangle=(a_1,b_1)^T$ and $|\psi_2 \rangle
 =(a_2,b_2)^T$, then a superposed state
 $ |\psi \rangle =( \alpha |\psi_1 \rangle + \beta|\psi_2 \rangle)$
corresponds to
\beq
z=\frac{\alpha a_1+ \beta a_2}{\alpha b_1 + \beta b_2},
\label{linear}
\eeq
due to the linearity of quantum mechanics and eq.(2). 

The inner product is given by
\beqa
\langle \psi_1|\psi_2 \rangle &=& a_1a_2+b_1^* b_2 \\ \nonumber
&=& \frac{|z_1||z_2|}{(|z_1|^2+1)
^{1/2}(|z_2|^2+1)^{1/2}}
\left( 1+\frac{1}{z_1^*z_2} \right ),
\eeqa
where $a_i (i=1,2)$ are non-negative real numbers, according to  the
definition in Eq.(\ref{definition}).

Now, let us show the relation between single qubit unitary
operations and conformal mappings on $\tilde{C}$.
An arbitrary unitary operation on the single qubit can be represented
by a unitary matrix on $C^2$
\beq
U=  e^{i\alpha}\left(
  \begin{array}{cc}
    c & d \\
    -d^* & c^*
  \end{array}
\right),
\label{U}
\eeq
which is a member of $SU(2)$ group multiplied by
$e^{i\alpha}$.
Here $|c|^2+|d|^2=1$ and
$c^*$  is the complex conjugate of $c$.
This  can be  also written in the form
\beq
U= e^{i \alpha} R_{\hat{n}}(\beta)
\eeq
with a rotation matrix $R_{\hat{n}}$ of the Bloch sphere by an
angle $\beta$  about an axis  $\hat n$.
On the other hand,   a  rotation group of the Bloch sphere
corresponds  to a special subset of the  M\"{o}bius transformation on the
extended complex plane
,which is given by
\beq
z'=f_U(z)\equiv \frac{ c' z + d'} {-d'^* z+ c'^*},
\label{fu}
\eeq where
$ |c'|^2 +  |d'|^2 = 1$.
One can define a matrix of coefficients of Eq.(\ref{fu});

\beq
U_f \equiv  \left(
  \begin{array}{cc}
    c' & d' \\
    -d'^* & c'^*
  \end{array}
\right)
,
\eeq
which is unitary and has determinant one.
We  find a relation between $U$ and $U_f$
as follows.
Under the unitary operation(Eq.(\ref{U}))
  the qubit $|\psi \rangle=(a,b)^T$
is transformed to $e^{i\alpha }( c a+ d b, -d^* a + c^* b)^T$, which
corresponds to
\beq
z'=\frac{ c a+db}{-d^* a + c^* b}
\label{z2}
\eeq
on $\tilde{C}$
according to  Eq.(\ref{z1}).
If  $|\psi\rangle =|1\rangle$ (i.e., $a=0,b=1$), then
  $ z=P(|\psi\rangle)=0$
  and $z'=f_U(0)= d'/c'^*$.
Comparing this with Eq.(\ref{z2}), we  find that
$d/c^*=d'/{c'}^*$.
From this we obtain $|d|^2/|c|^2=1/|c|^2-1=1/|c'|^2-1$,
which gives $c=e^{i\gamma} c'$,$ d=e^{-i\gamma} d'$ with
a constant $\gamma$.
 Similar argument for $P( [|0\rangle\pm |1\rangle]/\sqrt{2} )$
gives
\beq
\frac{-c+d}{-d^*+c^*}=\frac{-c'+d'}{-d'^*+c'^*},
\eeq
 which  demands $\gamma=0$ and finally
\beq
U= e^{i\alpha}U_f.
\label{relation}
\eeq
  This is our main result.
It is worthy to comment that $\alpha U_f$ with a constant $\alpha\in C, \alpha
\neq 0$
defines the same mapping $P$.
With this relation between unitary matrix $U$ and the matrix of
coefficients of the M\"{o}bius transformation $U_f$, one can
easily determine the  conformal mapping $f_U(z)$
  corresponding to the given single qubit unitary operation $U$
using  Eq.(\ref{fu}) and Eq.(\ref{relation}).
For example, the Hadamard gate
\beq
U=H=\frac{i}{\sqrt{2}}
 \left(
  \begin{array}{cc}
    1 & 1 \\
    1 & -1
  \end{array}
\right)
\eeq
on the Hilbert space  corresponds to the conformal mapping
\beq
z'=f_H(z)=\frac{ z + 1} { z-1 }
\eeq
on $\tilde{C}$.
This  can be checked by considering
the transformation of $P(|1\rangle)$ under the function  $f_{H}$;
\beqa
f_{H}(P(|1\rangle))&=&-1
=P \left(\frac{|0\rangle - |1\rangle}{\sqrt{2}}\right)\\ \nonumber
&=&P \left(i\frac{|0\rangle - |1\rangle}{\sqrt{2}}\right),
\eeqa
which implies $H |1\rangle =
i\frac{|0\rangle - |1\rangle}{\sqrt{2}}$ as desired.
Remember that there is a redundancy up
to a phase factor in single qubit states in this paper.
Table 1. shows the relationsphip between some single qubit
gates and their corresponding   M\"{o}bius transformations.

\vspace{1cm}
\begin{tabular}{|c|c|c|c|c|} \hline
Gate$(U)$& NOT(X) & Identity& Z  & Hadamard \\ \hline
$f_U(z)$ & $\frac{1}{z}$& $z$ & $-z$ & $\frac{ z + 1} { z-1 }$ \\ \hline
\end{tabular}
\vspace{0.5cm} \\
Table 1. Correspondence between single qubit unitary gates and
 M\"{o}bius transformations. Here $X$ and $Z$ are the Pauli operators.
\newline

Using the table one can also check operator product identities.
For example,
\beqa
f_{HZH}(z)&=&f_{HZ} \left( \frac{z+1}{z-1}\right)
=f_H \left( -\frac{z+1}{z-1}\right) \\ \nonumber
&=&\frac{1}{z}=f_X(z),
\eeqa
which implies $HZH=X$ as might be expected.
Note again that  operator relations also hold
 up to a global phase factor.

Let us discuss the time evolution of the states $|\psi(t)\rangle$
\beq
|\psi(t)\rangle=U(t)|\psi(0)\rangle
\eeq
with a time evolution operator $U(t)$.
For a constant Hamiltonian the Bloch vector  precesses  around an
axis given by the Hamiltonian.
If we denote the Hamiltonian $\hat{H}$ as $(H_0+
\sum_i \sigma_i H_i )$,
where $\sigma_i (i=1,2,3)$ are the
 Pauli matrices and   $H_0$ and   $H_i$
 parameterize the Hamiltonian, then

\beq
U(t)=e^{-i\hat{H}t/\hbar}=
 \left[
  \begin{array}{cc}
    c-iH_3s & -is(H_1-iH_2) \\
    -is(H_1+iH_2) & c+isH_3
  \end{array}
\right].
\eeq
Here $c\equiv cos(\theta),s\equiv sin(\theta)$
and $ \theta\equiv |\vec{H}|t/\hbar$.

This implies
\beq
z'=
 \frac{ (c-iH_3s)z -is(H_1-iH_2)}{-is(H_1+iH_2)z + c+isH_3}.
\eeq
Since any circle not containing the north pole of the Bloch sphere
is projected to a circle on $\tilde{C}$ under $P$, the precession
of the Bloch vector
 makes a circle  on $\tilde{C}$.

 There is
a possibility that utilizing fertility of the conformal theory,
 one can easily calculate complicated
quantities in the quantum computation theory and two-level
quantum systems.
For example, if the system has a conformal symmetry, then one might be
 able to calculate
 correlations using the conformal field theory\cite{cft}.

In summary,
identifying the Bloch sphere representing a single qubit state
with the Riemann , we  obtain
relations between  single qubit unitary
operations and M\"{o}bius transformations on the extended complex plane.
Using the relations we can  improve  our mathematical
and geometrical understanding of the qubit systems and
two-level quantum systems.

\vskip 1cm
This work
is supported in part by BK21.

\end{document}